\documentclass[11pt,preprint]{aastex}

\def\dd{{\rm d}}
\def\dg{{\rm o}}
\def\be{\begin{equation}}
\def\ee{\end{equation}}

\def\tp{t_p}

\def\gav{\bar{g}}
\def\Mmin{M_{\rm min}}
\def\Mtrue{M_{\rm true}}
\def\Mtrial{M_{\rm trial}}
\def\amin{a_{\rm min}}
\def\amax{a_{\rm max}}
\def\mtrue{m_{\rm true}}
\def\qtrue{q_{\rm true}}
\def\btrue{b_{\rm true}}

\def\Var{{\rm Var}}
\def\rp{r_1}
\def\rap{r_2}
\def\sp{s_1}
\def\sap{s_2}

\newbox\grsign \setbox\grsign=\hbox{$>$} \newdimen\grdimen \grdimen=\ht\grsign
\newbox\simlessbox \newbox\simgreatbox \newbox\simpropbox
\setbox\simgreatbox=\hbox{\raise.5ex\hbox{$>$}\llap
     {\lower.5ex\hbox{$\sim$}}}\ht1=\grdimen\dp1=0pt
\setbox\simlessbox=\hbox{\raise.5ex\hbox{$<$}\llap
     {\lower.5ex\hbox{$\sim$}}}\ht2=\grdimen\dp2=0pt
\setbox\simpropbox=\hbox{\raise.5ex\hbox{$\propto$}\llap
     {\lower.5ex\hbox{$\sim$}}}\ht2=\grdimen\dp2=0pt

\begin{document}

\title{Orbital roulette: a new method of 
   gravity estimation from observed motions}

\author{Andrei M. Beloborodov\altaffilmark{1,2}, Yuri Levin\altaffilmark{3}} 

\altaffiltext{1}{Physics Department, Columbia University, 538  West 120th 
Street New York, NY 10027}

\altaffiltext{2}{Astro-Space Center of Lebedev Physical 
Institute, Profsojuznaja 84/32, Moscow 117810, Russia} 

\altaffiltext{3}{Canadian Institute for Theoretical Astrophysics,
%University of Toronto, 
60 St. George Street, Toronto, ON M5S 3H8, Canada}

\begin{abstract}
The traditional way of estimating the gravitational field 
from observed motions of test objects is based on
the virial relation between their kinetic and potential energy.  
We find a more efficient method. It is based on the natural 
presumption that the objects are observed at a random moment of time 
and therefore have random orbital time phases. The proposed 
estimator, which we call ``orbital roulette'', checks the randomness
of the phases. The method has the following advantages: 
(1) It estimates accurately Keplerian (point-mass) potentials as well as 
non-Keplerian potentials where the unknown gravitating mass is 
distributed in space.
(2) It is a complete statistical estimator: it checks a trial 
potential and accepts it or rules it out with a certain significance level; 
the best-fit measurement is thus supplemented with error bars at any 
confidence level. 
(3) It needs no {\it \`a priori} assumptions about the distribution 
of orbital parameters of the test bodies. 
We test our estimator with Monte-Carlo-generated motions and demonstrate 
its efficiency. Useful applications include the Galactic Center, 
dark-matter halo of the Galaxy, and clusters of stars or galaxies.

\end{abstract}


%##############################################################################

\section{Introduction}

Estimation of a central point mass $M$ from measured positions and 
velocities of its $N$ satellites is a classic problem of astronomy
which applies to various gravitating systems.
More generally, the gravitating mass is distributed in space
with density $\rho({\bf r})$, so that $N$ test bodies move in an unknown 
potential $\Phi({\bf r})$, and astronomers estimate $\Phi$ from the 
observed motions.
In many cases the orbital periods of the test bodies are much longer 
than the age of modern astronomy and therefore only instantaneous 
positions and velocities are available rather than the full orbits of 
the bodies. This does not allow one to find the exact $M$ even in a 
point-mass problem. Instead, statistical methods are used to obtain
an approximate estimate.

Customary methods of mass estimation stem from the virial relation 
between the mean potential energy and mean kinetic energy. It enables an 
estimation of $\Phi$ from the observed kinetic energy of test bodies. 
One notes, however, that in a {\it finite} system of $N$ test bodies, the 
{\it instantaneous} relation between their potential and kinetic energies 
deviates from the virial relation. The variance of such deviations is 
unknown to the observer because it depends on the unknown orbital parameters 
of the bodies. For this reason, the observer does not possess a well defined 
error of the estimated mass.

The virial theorem itself is a statistical statement about the potential
and kinetic energies that is based on the presumption that {\it the 
observed bodies have random orbital time phases}.\footnote{The random-phase 
condition is satisfied in many astronomical systems including stellar 
clusters and clusters of galaxies (in planetary systems it may not hold if 
planets interact with each other and are locked in resonances). The virial 
relation is evidently not valid if the motions of the test bodies are 
correlated and the time of observation is chosen so that, for example, all 
bodies are near their pericenters. We do not consider the case of correlated 
test bodies here.}
Thus, the virial relation is not the original information we possess but a 
derivative of the random-phase principle. The latter simply states that 
the time we point our telescope to the system is random from the point of 
view of each body. 
One can make use of this basic principle directly, without
invoking the virial theorem. In this paper, we follow this approach and
develop a new method of potential estimation which we call orbital roulette 
(because the true orbital phase obeys the same statistics as a fair roulette, 
see \S~3). We show that this method is principally better and in practice 
more efficient than the virial estimator.

In many applications, the estimation of potential $\Phi$ is further 
complicated by the lack of full 3D information on the positions ${\bf r}_i$ 
and velocities ${\bf v}_i$ of test bodies: often only projections on the 
sky or line-of-sight components are measured. A projected version of 
orbital roulette will be addressed in a future paper. 
Here we develop the basic method in its full 3D version and discuss
the astronomical data sets to which it is immediately applicable.

The idea of our method is as follows. Consider a point-mass problem with
Keplerian potential $\Phi=-GM/r$ and suppose one tries to infer the 
central mass $M$ from observed motions of its satellites.
For any trial mass $\Mtrial$, the observed positions 
${\bf r}_i$ and velocities ${\bf v}_i$ of the satellites uniquely 
determine their orbits. Thus, for any $\Mtrial$, one can find 
the current orbital phases of the satellites. If $\Mtrial$ is smaller than 
the true mass $\Mtrue$, the orbital phases will be found near the 
pericenters, and if $\Mtrial>\Mtrue$ then the phases will cluster near 
the apocenters. Only for $\Mtrial=\Mtrue$ is the orbital roulette unbiased
and the calculated phases are randomly distributed between the 
pericenter and the apocenter. One can therefore figure out $\Mtrue$ by 
checking the randomness of the inferred phases. 
In \S~3, we put this idea on a quantitative basis and show how the lower 
and upper bounds on $M$ are obtained at a given confidence level. 

In \S~4 we extend the method to non-Keplerian potentials that have more 
than one unknown parameter and refine the roulette principle: the inferred 
orbital phases must be consistent with random numbers {\it and} uncorrelated 
with the inferred integrals of motion. This refinement is important for 
estimation of potentials created by distributed mass. 
 
One problem in the previously proposed methods was the need for 
{\it \`a priori} assumptions about the test bodies' population. For instance,
Little \& Tremaine (1987) and Kochanek (1996) used Bayesian analysis
to estimate the mass of our Galaxy. In order to make progress, they
needed some assumptions about the Galactic satellites' 
distribution function.\footnote{The Bayesian method also makes use of
the roulette principle because the satellite distribution 
function is assumed to depend only on the integrals of motion.
This form of the distribution function is justified by the strong Jeans 
theorem whose proof relies on the randomness of orbital phases 
(Appendix 4A in Binney \& Tremaine 1987).}
The roulette method makes no assumption about the orbits of test bodies. 
For a trial $\Phi({\bf r})$, it simply reconstructs the orbits and 
checks the obtained orbital phases. $\Phi({\bf r})$ is deemed a good estimate
at a confidence level $C$ if the phases are consistent with random 
distribution at the confidence level $C$.
This test extracts the maximum information 
on $\Phi({\bf r})$ one could possibly extract from data.

Any statistical estimator in general should be able to judge a trial 
gravitational potential $\Phi({\bf r})$ and accept it or 
rule out with a certain significance. Thus, the allowed hypotheses can 
be sorted out at a given confidence level.
The virial estimator does not work in this way: it gives 
only a best bet on the mass and a rough guess of its error based on the 
previous experience with numerical simulations of gravitating systems.
One of the advantages of the estimator we propose is its completeness in the 
sense that it enables derivation of the errors at any given confidence level.

The adequate statistical approach becomes especially valuable when one tries 
to discriminate between different models of $\Phi({\bf r})$ that have more 
than one parameter. Consider for example a set of trial models with two 
parameters $p_1$ and $p_2$. Which model is in best agreement with 
positions and velocities of the observed bodies? What $p_1$ and $p_2$ 
are allowed at the 90\% confidence level? The virial relation is not able 
to answer such questions. It gives instead a relation between 
$p_1$ and $p_2$ (and an approximate error of this relation). 
The roulette estimator turns out to be much more powerful: it constrains
both $p_1$ and $p_2$ at any chosen confidence level. We illustrate this 
in \S~4 with a simple halo model that has two parameters:
the halo size $b$ and mass $m$. In this example, the instantaneous 
positions and velocities of $N$ test bodies moving in the halo potential 
contain information on $m$ and $b$ which one would like to extract. 
The roulette method gives a best fit $(m_0,b_0)$ as well 
as confidence contours on the $(m,b)$ plane, and we test its 
efficiency with Monte-Carlo generated sets of bodies in a known potential
$(\mtrue,\btrue)$. We find that the standard deviation of  
$(m_0,b_0)$ from the true values is as small as 10-15\% for $N=32$.

The success of the roulette method is partially due to its ability to use
the full 3D information on the positions ${\bf r}_i$ and velocities 
${\bf v}_i$ of the test bodies. By contrast, the virial estimator uses only 
absolute values $r_i$ and $v_i$ ($i=1,...,N$), and thus the angles between 
${\bf v}_i$ and ${\bf r}_i$ are ignored. These angles are 
quite valuable as they contain
information on the eccentricities of the orbits. For instance, if ${\bf v}_i$ 
is parallel to ${\bf r}_i$ it is clear that the body is
on a radial (linear) orbit, while the virial estimator is not ``aware'' of 
that. The advantage of the roulette method in this respect becomes less
clear in applications where only projected components of ${\bf r}_i$ and
${\bf v}_i$ are known from observations. The method then needs to be
modified, which is deferred to a future paper. Already in its 3D version, 
the roulette estimator has useful applications that are briefly 
discussed in \S~5.

%##########################################################################

\section{Virial estimator}

In this section, we discuss briefly the performance of the virial estimator 
applied to $N$ test bodies orbiting a point mass $M$. It will be used later 
as a benchmark. The virial theorem states for any bound orbit 
\be
\label{eq:vir}
 <\frac{GM}{r}>=<v^2>,
\ee
where angle brackets signify a time average. The same relation holds 
for a population of $N\rightarrow\infty$ bodies 
observed at one moment of time if the average is taken 
over the population.\footnote{The virial relation~(\ref{eq:vir}) may be 
invalid if the population has orbits with strongly varying size $a$, so 
that $\amax/\amin\rightarrow \infty$ when $N\rightarrow\infty$.} 
For a set of observed positions $r_i$ and 
velocities $v_i$ one gets the virial mass
\be
\label{eq:vir1}
   M_v=\frac{\sum v_i^2}{G\sum 1/r_i}, \qquad i=1,...,N.
\ee
$M_v$ converges to $\Mtrue$ at $N\rightarrow\infty$, independently of the 
(unknown) orbital eccentricities of the bodies which makes the virial
relation tempting for mass estimations. The virial estimator has, however, 
the following drawbacks:

1. --- It does not involve any statistical analysis of the data. 
Instead of confidence intervals the estimator gives a direct formula for $M$ 
in terms of observed positions and velocities of the test bodies.
For a finite $N$, $M_v$ is not equal to the true mass, and its mean 
expectation and variation depend on the unknown orbital eccentricities of 
the bodies $e_i$. 

This is evident in the extreme case of $N=1$.
Then equation~(\ref{eq:vir1}) gives $M_v=G^{-1}v^2r$ where $v$ and $r$ are 
the velocity and radial position of the observed body.
The mean expectation for $M_v$ is $<M_v>=G^{-1}<v^2r>$ where the average
is now taken over the ensemble of random realizations of the data set. 
Each realization can be thought of as a snapshot taken at a random 
moment of time and $<v^2r>$ equals the time-averaged value of 
$v^2r$ for the true orbit (see Appendix), 
\be
\label{eq:v2r}
  <v^2r>=G\Mtrue\left(1-\frac{e^2}{2}\right).
\ee
The standard deviation $\Delta(v^2r)$ can also be calculated,
\be
   \frac{\Delta(v^2r)}{<v^2r>}=\frac{e}{\sqrt{2-e^2}}. 
\ee
Thus, both $<M_v>$ and $\Delta(M_v)$ depend on the unknown $e$
for $N=1$. 

A qualitatively similar dependence takes place for $N>1$, and 
therefore it is difficult to derive the error of $M_v$ without additional
assumptions about $e_i$.
The error would vanish for circular orbits ($e_i=0$): each body would be 
expected to give the same $M=v_i^2r_i/G$, $i=1,...,N$. In the case of 
large $e_i$ the error can be significant even if $N\gg 1$ and depends 
on the radial distribution of the bodies as we discuss next. 
 
2. --- Test bodies that are at larger radii contribute to the virial estimator
with a smaller weight and those at the smallest radii make the dominant 
contribution. If $N$ bodies are spread over a significant range of radii, 
the effective size of the sample is smaller than $N$, which leads to a
relatively large statistical error of the estimated mass.

For illustration, consider a population of orbits with random eccentricity
$0<e<1$ and semi-major axis $\amin<a<\amax$, and make a simple Monte-Carlo 
simulation. 
For each body $i=1,...,N$ we draw randomly $e_i$, $a_i$, and
the orbital time phase of the body. Thus we create a ``data set'' 
${\bf r}_i$, ${\bf v}_i$. We apply equation~(\ref{eq:vir1}) to the set
and obtain $M_v$, which we can compare with the true $M$. Repeating this 
for many randomly drawn data sets we can study the statistics of $M_v$.

Figure~1 shows the standard deviation of $M_v$ for $N=10$, 100, 1000. 
It depends on the ratio $\amax/\amin$: with increasing $\amax/\amin$ the 
statistical error of $M_v$ increases. This happens because the bodies at 
small $r$ (and correspondingly large $v$) dominate in 
equation~(\ref{eq:vir1}), and the data with large $r$ is practically lost. 

%%%%%%%%%%%%%%%%%%%%%%%%%%%%%%%%%%%%%%%%%%%%%%%%%%%%%%%%%%%%%%
\begin{figure}
\begin{center}
\plotone{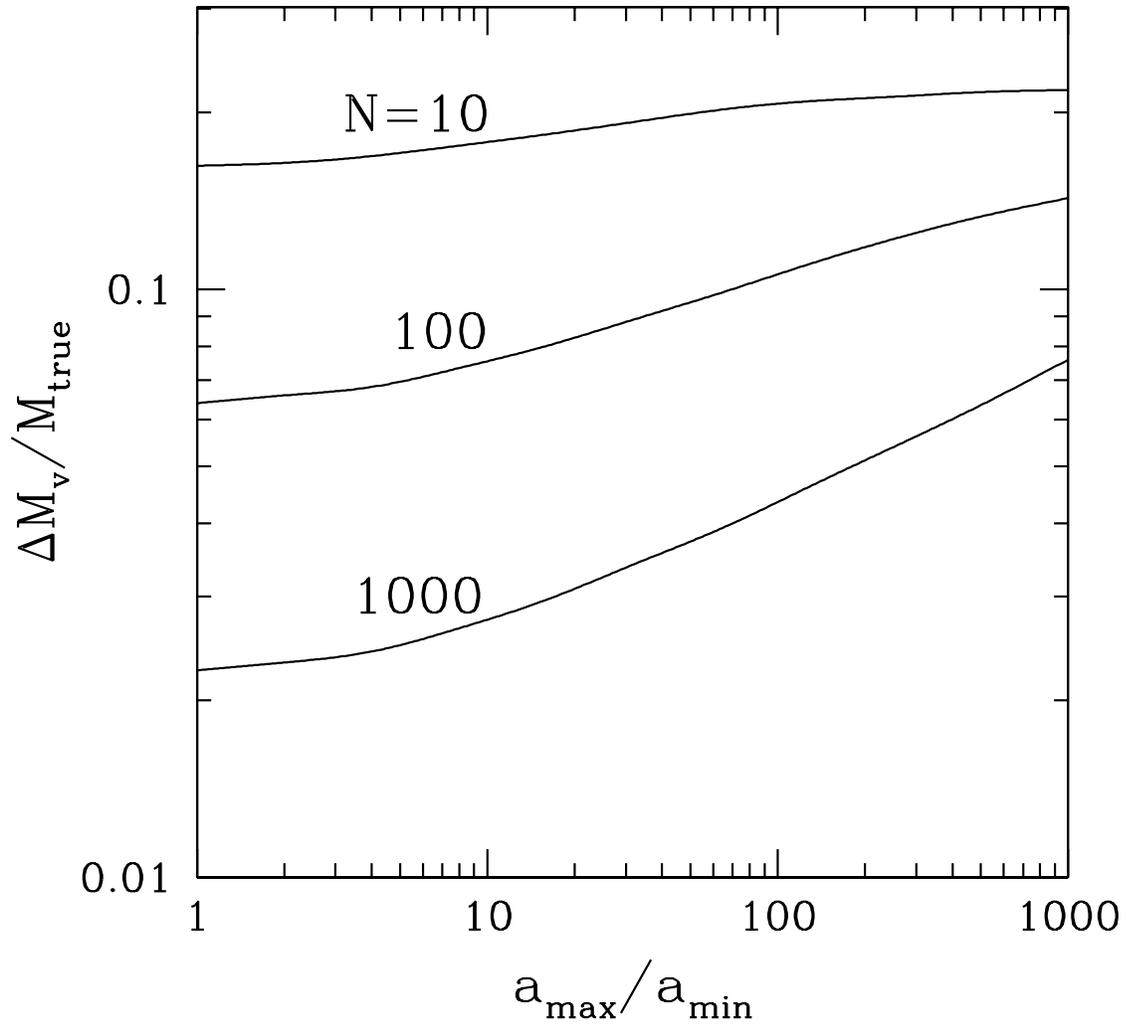}
\label{fig:vir}
%\epsfxsize=17cm
%\epsfysize=17cm
%\epsfbox{f1.eps}
\end{center}
\caption{ Standard deviation of the virial estimate $M_v$ for a point mass
with $N$ observed satellites. The orbits of test bodies are assumed to have 
random orbital eccentricities $0<e<1$ and semi-major axes $\amin<a<\amax$. 
}
\end{figure}
%%%%%%%%%%%%%%%%%%%%%%%%%%%%%%%%%%%%%%%%%%%%%%%%%%%%%%%%%%%%%%

To avoid this problem one could relate $M$ to $<v^2r>$
instead of equation~(\ref{eq:vir1}).
This relation, however, depends on $e_i$, even at $N\rightarrow\infty$. 
For example, for bodies on orbits with equal 
eccentricities $e_i=e$ one gets (see eq.~\ref{eq:v2r})
\be
\label{eq:BT}
   M=\frac{<v^2r>}{G(1-e^2/2)}.
\ee
A projected version of this estimator was proposed by Page (1952) 
who studied the case of circular orbits ($e=0$) and by Bahcall \& Tremaine 
(1981) for arbitrary $e$. The advantage of such estimates is that bodies
at different $r$ contribute equally, which gives a nice convergence
of the estimated $M$ with increasing $N$.  
Nevertheless, the fact that $e$ is unknown still impedes a precise 
estimate of $M$ even when the exact $<v^2r>$ is known ($N\rightarrow\infty$).

3. --- The virial estimator uses only absolute values $r_i$ and $v_i$ and 
ignores the angles between ${\bf v}_i$ and ${\bf r}_i$. 
\medskip

Finally, we note that the virial estimator is inefficient when applied
to non-Keplerian potentials. The virial relation for a general potential
is given by
\be
\sum_{i=1}^N \left({\bf v}_i^2+{\bf r}_i\cdot {\bf f}_i\right)=0,
\ee
where ${\bf f}_i$ are the gravitational accelerations of the test bodies.
It gives one condition and, if the potential has more than one 
unknown parameter, the relation is not able to determine them.

%##########################################################################

\section{Orbital roulette}

Suppose we have measured a current 3D position ${\bf r}$ and a velocity 
${\bf v}$ of a satellite orbiting a central massive object. 
What can we say about the central mass $M$ based on this ``snapshot''
information only? First of all, assuming that the satellite is on a bound 
orbit (its age is longer than the orbital period),
we get a lower bound $M_{\rm min}=v^2r/2G$.
Then we note that if $M$ is very large, $GM/r\gg v^2/2$, then the satellite 
has to be extremely close to its apocenter, and if $M$ is small, 
$M\rightarrow M_{\rm min}$, it has to be near the pericenter (Fig.~2). 
The probability of either extreme is low because the snapshot is taken 
at a random moment of time and, 
in most cases, the satellite should be somewhere between the pericenter
and apocenter. Thus, the requirement of a random orbital time phase 
can constrain $M$ from above and below at a given confidence level. 
If we have a snapshot of $N\gg 1$ satellites, the constraint becomes tight.
Below we quantify this constraint and develop the mass estimator based
on the random-phase principle. The simplest way to check the 
randomness is by comparing the mean phase with its expected value 
(\S~3.1) and a more sophisticated method checks the whole distribution 
of $N$ phases (\S~3.2).

%%%%%%%%%%%%%%%%%%%%%%%%%%%%%%%%%%%%%%%%%%%%%%%%%%%%%%%%%%%%%%
\begin{figure}
\begin{center}
\plotone{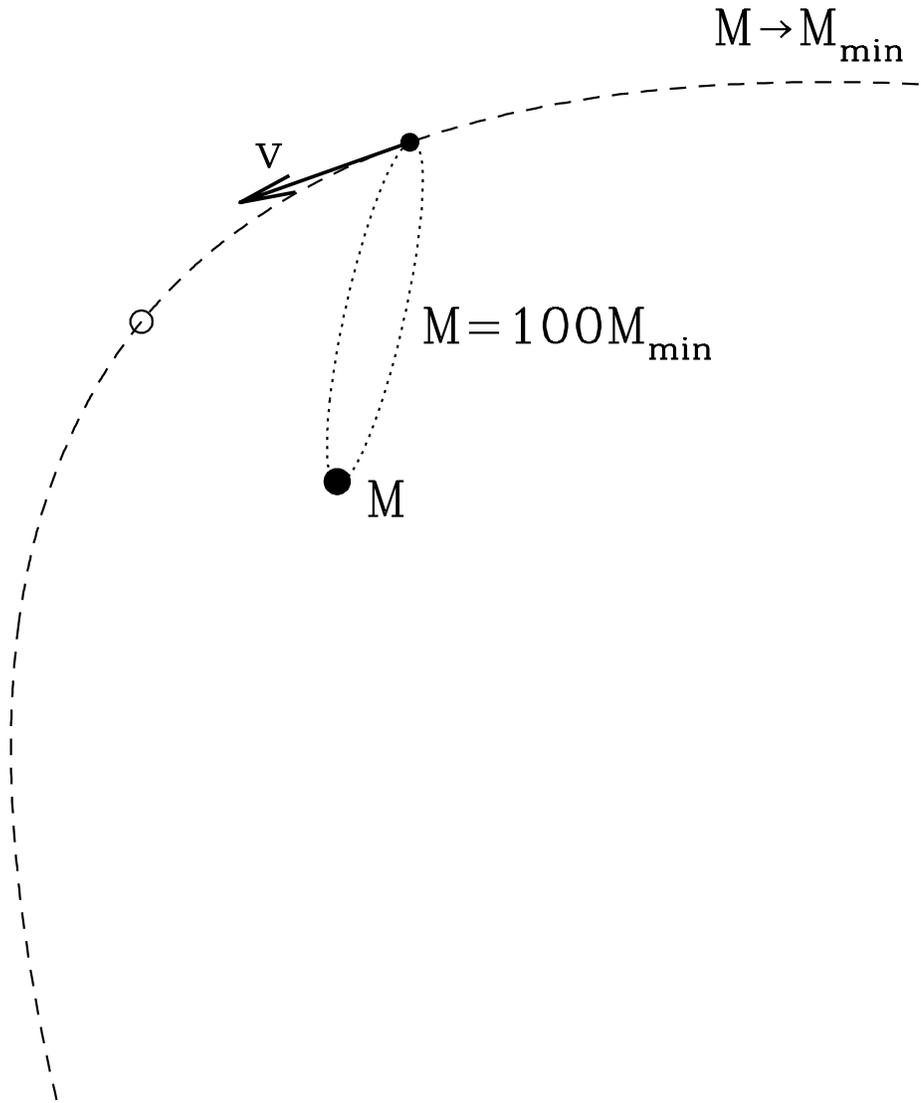}
%\label{fig:V}
%\epsfxsize=17cm
%\epsfysize=17cm
%\epsfbox{f1.eps}
\end{center}
\caption{Orbit reconstruction for a body with a given position and velocity 
vector ${\mathbf v}$. The reconstructed orbit depends on the assumed central
mass $M$. A small $M\rightarrow M_{\rm min}=rv^2/2G$ gives a large orbit and 
then the current position is much closer to the pericenter (shown by the open
circle) than a distant apocenter far outside the figure. A large $M$ 
($100M_{\rm min}$ in this example) gives a small orbit and places the body 
almost exactly at the apocenter. 
}
\end{figure}
%%%%%%%%%%%%%%%%%%%%%%%%%%%%%%%%%%%%%%%%%%%%%%%%%%%%%%%%%%%%%%

\subsection{Mean orbital phase}

Let us denote the unknown orbital period of a satellite by $T$ and
the time since last passage of the pericenter by $\tp$.
Since we know the current ${\bf r}$ and ${\bf v}$,
the whole orbit and $\tp$ are unambiguously calculated
for any given $\Mtrial$. For a known orbit,
we need a measure of how close the satellite is to the pericenter.
We define this measure as time separating the satellite from the nearest
passage of the pericenter (in the past or the future) and normalize it by 
$T/2$,
\be
\label{eq:g}
  g(\Mtrial)=\frac{1}{T/2}\left\{\begin{array}{ll}
  t_p & v_r>0, \\
  T-t_p & v_r<0, 
              \end{array}
       \right.
\ee
so that $g$ can take values between 0 and 1. This definition takes into 
account that the pericenter is closer in the past if the current radial 
component of velocity $v_r>0$, and in the future if $v_r<0$.
The limit $\Mtrial\rightarrow \Mmin$ gives $g\rightarrow 0$ (pericenter), 
and $\Mtrial\rightarrow\infty$ gives $g\rightarrow 1$ (apocenter). 
For the true $M$ and a random snapshot time, the expected $g$ obeys
Poisson statistics: it has a flat probability distribution between 0 and 1 
with the mean expectation value $<g>=1/2$ and the standard deviation
$\Delta g= 12^{-1/2}$.

Now suppose we have a snapshot of $N\gg 1$ satellites with measured 
${\bf r}_i$ and ${\bf v}_i$, $i=1,...,N$. For a given $\Mtrial$ we 
can calculate the orbit of each satellite and the corresponding 
$g_i(\Mtrial)$. Then we define the mean phase,
\begin{equation}
  \bar{g}(\Mtrial)=\frac{1}{N}\sum_{i=1}^N g_i(\Mtrial).
\end{equation}
Again, $\bar{g}\rightarrow 0$ for small $\Mtrial$ and 
$\bar{g}\rightarrow 1$ for large $\Mtrial$. By contrast, for the true $M$, the 
expected $\gav$ is described by a narrow probability distribution $f(\gav)$ 
which peaks at $1/2$. At $N\gg 1$, this distribution is Gaussian 
according to the central limit theorem,
\be
\label{eq:Gauss}
  f(\gav)=\frac{1}{\sqrt{2\pi}\sigma}
          \exp\left[-\frac{(\gav-1/2)^2}{2\sigma^2}\right], 
       \qquad \sigma=\frac{1}{\sqrt{12N}}.
\ee
So, one can find the gravitating mass by adjusting $M$ so that 
\be
\label{eq:roul}
   \bar{g}(M)=\frac{1}{2}\pm(12N)^{-1/2}.
\ee
The estimate is well defined since $\gav$ is a monotonic function of $M$.
The width $\sigma$ of the distribution~(\ref{eq:Gauss}) sets the error of 
the estimate. Expanding equation~(\ref{eq:roul}) near the best-fit point 
$\gav=0.5$, one gets the error in $M$,
\be
\label{eq:roul_er}
  \frac{\Delta M}{M}=(12N)^{-1/2}
                   \left[M\frac{\dd \gav}{\dd M}\right]^{-1}_{\gav=1/2}.
\ee 
$M(\dd\gav/\dd M)=O(1)$ for any large $N$ (variation of $M$ by a factor 
$\sim 2$ around the best-fit value induces a change of $\gav$ comparable
to $1/2$) and hence $\Delta M/M=O(N^{-1/2})$. The estimate of $M$ defined 
by equation~(\ref{eq:roul}) converges to $\Mtrue$ at $N\rightarrow\infty$.

The analysis of orbital phases $g_i$ is similar to testing a roulette 
wheel. For a fair roulette, the ball angular position on the wheel must be 
random. Suppose we have a reference position, e.g. Zero, 
and make $N$ experiments with the roulette. Then we get $N$ random angular 
deviations from Zero, $0\leq\alpha_i\leq 180^\dg$, $i=1,...,N$.
If the wheel is unbiased, one expects $\alpha_i$ to follow Poisson statistics 
with a mean value $\bar{\alpha_i}=[0.5\pm (12N)^{-1/2}]\times 180^\dg$. 
There is a full analogy with our problem as the magnitude 
$\bar{g}=\bar{\alpha}/180^\dg$ obeys the same probability distribution
(eq.~\ref{eq:Gauss}). The only difference is that the random variable of 
orbital roulette is the {\it time} phase of a test body rather than its
angular position.

Equation~(\ref{eq:roul}) gives a best-bet value of $M$ and an estimate 
of its error. We now aim to obtain a more complete solution to the problem: 
the allowed interval for $M$ at a given confidence level $C$. 
Mathematically, 
this is formulated as follows. For a given number $0<\xi<1/2$ we evaluate 
$M_+$ such that the probability of $\Mtrue>M_+$ equals $\xi$, and $M_-$ 
such that the probability of $\Mtrue<M_-$ equals $\xi$. The interval 
$M_-(\xi)<M<M_+(\xi)$ is the mass measurement at the confidence level 
$C=1-2\xi$.

Let us define cumulative distributions
\be
\label{eq:Ppm}
  P_-(g)=\int_0^{g} f(\gav)\dd\gav, \qquad
  P_+(g)=\int_{g}^1 f(\gav)\dd\gav.
\ee
$P_-+P_+=1$ for any $\Mtrial$. Then define $g_-(\xi)$ and $g_+(\xi)$ so that
\be
\label{eq:gpm}
  P_-(g_-)=\xi, \qquad P_+(g_+)=\xi.
\ee
Note that $g_\pm$ are unique functions of $\xi$ (they depend only on $N$ 
as a parameter) and the snapshot data ${\bf r}_i$ and ${\bf v}_i$ did 
not appear in the definitions~(\ref{eq:Ppm},\ref{eq:gpm}). The data 
determine $\gav(M)$ and appear in the final equation for $M_\pm(\xi)$,
\be
\label{eq:Mpm}
  \gav(M_-)=g_-, \qquad \gav(M_+)=g_+.
\ee

One can prove that $M_\pm$ defined in this way give the correct confidence 
interval for $M$ at the confidence level $C=1-2\xi$. Consider the ensemble of 
all possible snapshots of our test bodies taken at random moments of time. 
For each snapshot one has $M_-(\xi)$ defined in equation~(\ref{eq:Mpm}).
What is the fraction of ``bad'' snapshots with $\Mtrue<M_-(\xi)$, outside 
the confidence interval? For these snapshots, 
\be
  \gav_{\rm true}\equiv\gav(\Mtrue)<g_-(\xi),
\ee
[this is equivalent to $\Mtrue<M_-(\xi)$ because $\gav(M)$ is a 
monotonic function for any snapshot].
The probability to get a snapshot with $\gav_{\rm true}<g_-$ is 
$P_-(g_-)$, and it equals $\xi$ by definition of $g_-$. 
Hence, the probability of a snapshot with $\Mtrue<M_-(\xi)$ equals $\xi$. 
Analagously, one shows that the probability of a snapshot with 
$\Mtrue>M_+(\xi)$ equals $\xi$. 

$P_-(\gav)\ll 1$ signals $\Mtrial<\Mtrue$, and $P_+(\gav)\ll 1$ 
signals $\Mtrial>\Mtrue$. The best bet for $M$ is defined by 
$P_-(M)=P_+(M)=1/2$, which is equivalent to $\gav(M)=0.5$. 
The best bet is supplemented with the confidence intervals $(M_-,M_+)$ 
at any confidence level, which also gives an explicit probability 
distribution of $M$,
\be
  p(M)=\frac{\dd P_-}{\dd M}=-\frac{\dd P_+}{\dd M}
      =\frac{\dd \gav}{\dd M} f(\gav[M]).
\ee
This probability distribution is the measurement of $M$ with the mean-phase
method.

\subsection{Cumulative distribution of orbital phases}

When the trial mass is close to $\Mtrue$, the inferred phases $g_i$ 
should be consistent with a Poisson distribution, which can be used as 
a test for $\Mtrial$. Such a test has the advantage of using the whole 
distribution of $g_i$ rather than just the average $\gav$.
Consider, for example, $g_i$ half of which
equal 0 and the other half equal 1. This is clearly inconsistent with 
a Poisson distribution, yet $\gav$ has the right value of 1/2  
and the mean-phase method will not notice the inconsistency.

Testing a distribution for consistency with an expected/model distribution 
(null hypothesis) is a standard problem of mathematical statistics. 
It is done by comparing the corresponding cumulative distributions.
(Kolmogorov 1941).
The cumulative distribution of a given set $g_i$ ($i=1,...,N$) is a stepped 
function, $0\leq F(g)\leq 1$, that increases by $1/N$ at each $g_i$.
The null hypothesis in our case is a Poisson process, which gives the mean 
expectation $<F(g)>=g$ at any $g$ (and for any $N$). Random realizations
of the Poisson set $g_i$ have $F(g)$ fluctuating around $<F(g)>$ with a 
binomial distribution,
\begin{eqnarray}
\label{eq:Pois}
  p[F(g)]=\frac{N!}{(FN)!(N-FN)!}g^{FN}(1-g)^{N-FN}, \\
  <F(g)>=g, \qquad \Var[F(g)]=\Delta^2[F(g)]=\frac{g(1-g)}{N},
\label{eq:Pois1}
\end{eqnarray}
where $\Var[X]$ and $\Delta[X]$ denote variance and standard deviation of 
$X$. The unknown mass $M$ should be adjusted so that the cumulative 
distribution of $g_i(M)$ is consistent with a Poisson process.

One then needs a measure of the difference between the obtained 
$F(g)$ (which depends on $M$ as a parameter) and the mean expectation
$<F(g)>=g$. The simplest measure would be the Kolmogorov-Smirnov test: 
one defines $V_{\rm KS}=\max |F(g)-g|$, where the maximum is searched 
over the whole interval $0\leq g\leq 1$ (Kolmogorov 1941).
The probability distribution of $V_{\rm KS}$ is known for a true  
Poisson process (e.g., Press et al. 1992) and getting the actual 
$V_{\rm KS}$ far in the tail of this distribution signals a low 
probability of consistency with the null hypothesis.

The Kolmogorov-Smirnov measure is most sensitive to the values of $F(g)$ 
at median $g\sim 0.5$ where $\Var[F(g)]$ is largest (see eq.~\ref{eq:Pois1}).
In our case, a measure sensitive at the tails $g\approx 0$ and $g\approx 1$ 
would be preferable because, when $\Mtrial$ deviates from $\Mtrue$, $g_i$ 
cluster near 0 or 1. We therefore use a different measure
\be
\label{eq:V}
  V=\int_0^1\frac{[F(g)-<F(g)>]^2}{\Var[F(g)]}\,\dd g
   =N\int_0^1\frac{[F(g)-g]^2}{g(1-g)}\,\dd g.
\ee
It gives more weight to the tails and is known as the Anderson-Darling 
test (Anderson \& Darling 1952).  
It can be viewed as $\chi^2$ for a fit of $F(g)$ by $F=g$.
Equation~(\ref{eq:V}) becomes the usual formula for $\chi^2$ if the 
integral is replaced by a finite sum over $\dd g=1/K$ with 
$K\rightarrow\infty$. 

The probability distribution of $V$ for a true Poisson process
can be calculated numerically by the Monte-Carlo technique. We randomly 
generate many Poisson sets $g_i$ ($i=1,...,N$), calculate $V$
for each set, and find the histogram of $V$. The found probability 
distribution $p(V)$ is shown in Figure~3. 
It does not depend on $N$ as 
long as $N\gg 1$. At $V>2$ ($p<0.1$) the distribution is very well 
approximated by a simple empirical formula,
\be
\label{eq:pV}
 p(V)=10^{-V/2},  \qquad V>2.
\ee

Having $g_i(\Mtrial)$ and their $V$, one can quantify the inconsistency of 
the trial with a Poisson process. The significance level of inconsistency,
\be
  \xi(V)=\int_{V}^\infty p(V^\prime)\dd V^\prime, 
\ee
is the probability that $V^\prime$ computed for a random Poisson set 
$g_i^\prime$ is greater than $V$. A low $\xi$ means that it is unlikely to 
draw occasionally a Poisson set $g_i^\prime$ with $V^\prime$ as large as 
our $V$, and so our $\Mtrial$ is rejected at the significance level $\xi$. 

For $V(M)>2$, one can use equation~(\ref{eq:pV}) to get a simple explicit
formula
\be
\label{eq:xi}
  \xi(V)=\frac{2}{\ln 10}10^{-V/2}, \qquad V>2.
\ee
All $\Mtrial$ that give $V(\Mtrial)>-2\log_{10}[(\xi/2)\ln 10])$ are 
rejected at the significance level $\xi$. 

The function $V(\Mtrial)$ may be non-monotonic (especially near its minimum), 
and a number $k\geq 2$ of mass intervals might be rejected. 
$\Mtrue$ belongs to the remaining regions at the confidence level $C=1-k\xi$.
In practice, however, our Monte-Carlo simulations show that the case $k>2$ 
rarely occurs for $\xi$ of practical interest, and we count 
only two rejected intervals that cover the tails of low mass $(\Mmin,M_-)$ 
and high mass $(M_+,\infty)$. $\Mtrue$ belongs to the interval $(M_-,M_+)$ 
at confidence level $C=1-2\xi$.

%%%%%%%%%%%%%%%%%%%%%%%%%%%%%%%%%%%%%%%%%%%%%%%%%%%%%%%%%%%%%%
\begin{figure}
\begin{center}
\plotone{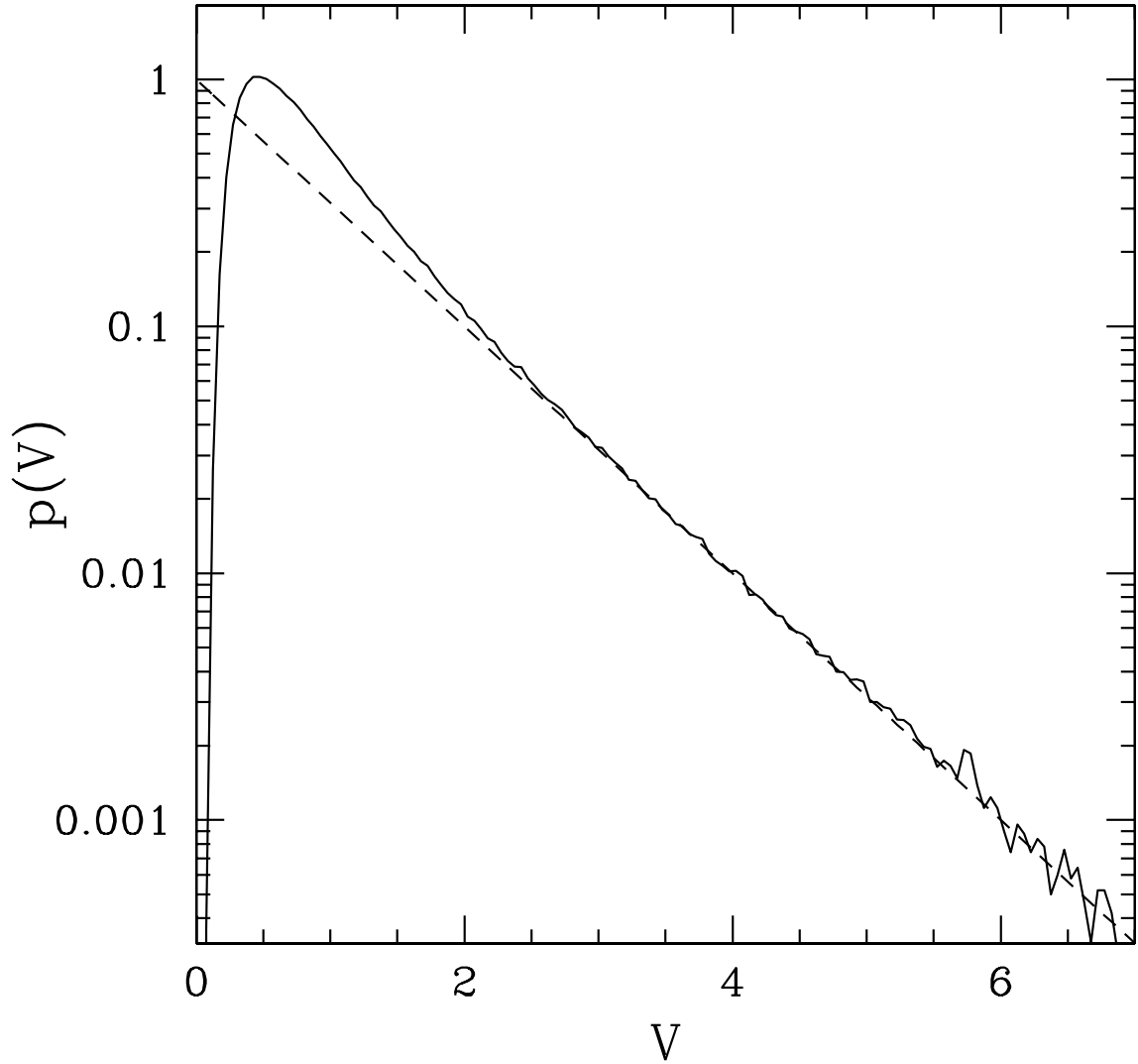}
%\label{fig:V}
%\epsfxsize=17cm
%\epsfysize=17cm
%\epsfbox{f1.eps}
\end{center}
\caption{Distribution of $V$ (eq.~\ref{eq:V}) for a true Poisson process 
with $N\gg 1$. The solid curve shows the result of numerical (Monte-Carlo)
calculation. The dashed line shows the approximation
$p(V)=10^{-V/2}$, which is valid at $V>2$ ($p<0.1$).
}
\end{figure}
%%%%%%%%%%%%%%%%%%%%%%%%%%%%%%%%%%%%%%%%%%%%%%%%%%%%%%%%%%%%%%

\subsection{Practical algorithm}

We summarize now the practical steps to derive the central point mass $M$ 
from data. Given a measured position ${\bf r}$ and velocity ${\bf v}$ of 
a test body, one first determines its orbital phase $g$ as a function of 
$M$. An explicit formula for $g$ is derived in Appendix~A,
\be
\label{eq:g_}
  g=\frac{1}{\pi}\left(2\psi-e\sin 2\psi\right), 
\ee
where 
\be
\label{eq:psi_}
  \cos 2\psi=\frac{1}{e}\left(1+\frac{2Er}{GM}\right),
\ee
and the orbital energy $E$ and eccentricity $e$ are expressed in terms of 
${\bf r}$, ${\bf v}$, and unknown $M$ in Appendix~A 
(eqs.~\ref{eq:E},\ref{eq:ecc}). Equation~(\ref{eq:g_}) gives $g_i(M)$
for each test body $i=1,...,N$. 

The mean-phase variant of the roulette estimator is simplest to apply.
It defines the best-fit $M$ by equation $\gav(M)\equiv\sum g_i/N=1/2$. 
Since $\gav$ increases monotonically with $M$, only one solution to
equation $\gav(M)=1/2$ can exist. 
The solution is easily found with the bisection method. 
The initial interval $(\Mmin,M_{\rm max})$ for bisection can be chosen as 
follows. $\Mmin$ is given by the condition that all $N$ orbits are bound:
$\Mmin=\max(v_i^2/2Gr_i)$. $M_{\rm max}$ should be taken sufficiently large, 
so that $\gav(M_{\rm max})>1/2$. Note that no solution exists if 
$\gav(\Mmin)>1/2$.
It may happen in rare cases where the observed bodies accidentally cluster 
near their true apocenters and $\gav(\Mtrue)$ is large.
Such snapshots may have $\gav(\Mmin)>1/2$, and then
the best-fit mass is not defined by the mean-phase estimator.
Only confidence intervals should be considered in this case.
  
The confidence intervals provide the most useful information.
They show the allowed interval for $M$ at a given confidence level 
$C=1-2\xi$ where $\xi$ can be chosen at any small value. In practice,
measurements at the 90\% confidence level are often considered, which 
corresponds to $\xi=0.05$. Confidence intervals $(M_-,M_+)$ derived with 
the mean-phase estimator are given by equation~(\ref{eq:Mpm}) where 
$g_\pm(\xi)$ are defined by equations~(\ref{eq:Ppm},\ref{eq:gpm}). 
Note that $f(\gav)$ is
precisely Gaussian only for $N\gg 1$, and at small $N$ one should use a
different formula: the convolution of $N$ constant functions $f_1(g)=1$,
$0<g<1$ (the convolution approaches Gaussian quickly with increasing
$N$; e.g. at $N=5$ the Gaussian approximation is already good). 

The test for the cumulative distribution of orbital phases is slightly 
more complicated, however, it has the advantage of utilizing all the 
available data. For any trial mass $M$ one can calculate the 
Anderson-Darling $V(M)$ (eq.~\ref{eq:V}), which can be written as
\be
\label{eq:V_}
 V=N\sum_{i=0}^N\left[\left(\frac{i}{N}\right)^2\ln\frac{g_{i+1}}{g_i}
   -\left(1-\frac{i}{N}\right)^2\ln\left(\frac{1-g_{i+1}}{1-g_i}\right)
   +g_i-g_{i+1}\right].
\ee
Formal $g_0\equiv 0$ and $g_{N+1}\equiv 1$ are used in this expression.
The best-fit $M$ is such that $V(M)$ is minimal. This minimum can be found 
numerically.
Alternatively, one solves the algebraic equation $\dd V/\dd M=0$, 
which can be written down analytically. The derivative $\dd V/\dd M$ is 
\be
\label{eq:dVdM}
  \frac{\dd V}{\dd M}=\frac{1}{N^2}\sum_{i=1}^N\frac{\dd g_i}{\dd M}
    \left[\frac{2(N-i)+1}{1-g_i}-\frac{(2i-1)}{g_i}\right],
\ee
and the expression for $\dd g/\dd M$ is derived in Appendix~A. This gives an 
explicit analytical form to equation $\frac{\dd V}{\dd M}(M)=0$. The 
equation may, however, have many roots, i.e., $V(M)$ may have many local 
minima and maxima around the global minimum. Therefore in practice it is 
easier to find the minimum numerically.

Once $V(M)$ is evaluated, one should find the intervals of $M$ that
can be rejected at a given significance level $\xi$ as explained in the end 
of \S~3.2. If $\xi<0.1$ (i.e. the estimate is done on a confidence 
level $C=1-2\xi>80$\%), the analytical approximation to $V(\xi)$ can
be used. Then the boundaries of the rejected intervals are solutions of 
the equation $V(M)=-2\log_{10}[(\xi/2)\ln 10]$.

\subsection{Handling data uncertainties}

The described method is easy to apply to a snapshot of $N$ bodies with
precisely measured ${\bf r}_i$ and ${\bf v}_i$. Real data, however, have 
uncertainties. Suppose a measurement gives
3D distributions $p({\bf v}_i)$ and $p({\bf r}_i)$
rather than numbers ${\bf v}_i$ and ${\bf r}_i$; for example,
they may be Gaussian distributions whose widths represent
the measurement errors of ${\bf r}_i$ and ${\bf v}_i$.
Then, for a given $\Mtrial$, the orbital parameters of the bodies 
are also described by probability distributions. As a result one obtains
$N$ probability distributions $p_i(g_i)$ and $p(\gav)$ instead of numbers 
$g_i$ and $\gav$. 

The $p(\bar{g})$ found for $\Mtrial$ should be compared with the 
expected distribution $f(\gav)$ (eq.~\ref{eq:Gauss}) and their 
consistency should be evaluated. This can be done using the
Kolmogorov-Smirnov or Anderson-Darling tests. 
In practice, it is easier to use the Monte-Carlo technique 
that incorporates the data uncertainties directly in the roulette test 
for $\Mtrial$. 
In each Monte-Carlo event, one draws randomly ${\bf v}_i$ and ${\bf r}_i$ 
from the measured distributions $p({\bf v}_i)$ and $p({\bf r}_i)$,  
calculates the orbits, and finds the mean orbital phase $\bar{g}$ of the 
$N$ satellites. Then one compares $\bar{g}$ with $\bar{g}^\prime$ 
for $N$ randomly drawn numbers $0<g_i^\prime<1$.
Repeating this comparison for many Monte-Carlo ${\bf v}_i$, ${\bf r}_i$ 
and $\bar{g}^\prime$, one finds the probability $P_-(\Mtrial)$ that
$\bar{g}>\bar{g}^\prime$ and the corresponding $P_+(\Mtrial)=1-P_-$. 
Thus, one gets the same estimator as the one described in \S~3.1 but now 
it allows for the data uncertainties. The best-fit mass is found from the 
condition $P_-=P_+=1/2$ and the confidence intervals are found from 
$P_-=P_+=\xi$.
In the case of large data uncertainties, the functions $P_\pm(\Mtrial)$ 
are modified and the range of $M$ consistent with the data becomes larger. 

The mass estimator based on the Anderson-Darling test of the cumulative 
distribution of $g_i$ is modified in a similar way to incorporate the 
data uncertainties.

In a separate paper on the Galactic Center, we present an analysis of a
real data set and demonstrate how the measurement errors are taken into 
account.

\subsection{Numerical tests}

We now check the performance of the roulette estimator by direct Monte-Carlo 
simulations. With this purpose, we have developed a numerical code that 
calculates the best-fit mass and confidence intervals with both mean-phase 
and cumulative-distribution variants of the estimator.

First of all we check that the confidence intervals are correct. 
We take $N$ bodies orbiting a central mass 
$\Mtrue$ with randomly chosen eccentricities and semi-major axes. 
Each body is taken at a random moment of time. This gives us a simulated 
data set to which we apply the estimator and obtain the 90\% confidence 
intervals $(M_-,M_+)$. We repeat this procedure for many generated data 
sets and count the number of cases where $\Mtrue$ is outside our 
confidence interval. These cases should make 10\% and we have checked
that this is indeed so.

Then we compare the roulette method with the virial estimator.
Since the virial estimator does not give confidence intervals, we 
have to restrict the comparison to the best-fit values of $M$.
We denote them by $M_r$ and $M_v$ for the roulette and virial estimators,
respectively. Computer-generated random data sets allow us to study the 
statistics of $M_r$ and $M_v$.

The advantage of the roulette estimator becomes significant when the 
generated orbits have significantly different sizes. For illustration, 
we draw random semi-major axes of the orbits from an interval $0<a<\amax$.
We also draw a random eccentricity $0<e<1$ for each orbit. 
Each generated data set of $N$ bodies is analyzed with three methods:
(1) virial, (2) roulette mean-phase test, and (3) roulette 
cumulative-distribution test.
The results of analysis of $10^5$ data sets are shown in Figure~4. 
We observe that for $N\ge 4$ the roulette method gives a more precise 
estimate of $M$ than the virial estimator and shows a much faster
convergence to $\Mtrue$ with increasing $N$. 
We also observe that the mean-phase estimator works practically as well as 
the more complete analysis of the cumulative distribution. 

The roulette method is illustrated in Figure~5 for six randomly selected
data sets with $N=10$. The best-fits and 90\% confidence intervals obtained 
with the two roulette estimators strongly correlate with each other.

%%%%%%%%%%%%%%%%%%%%%%%%%%%%%%%%%%%%%%%%%%%%%%%%%%%%%%%%%%%%%%
\begin{figure}
\begin{center}
\plotone{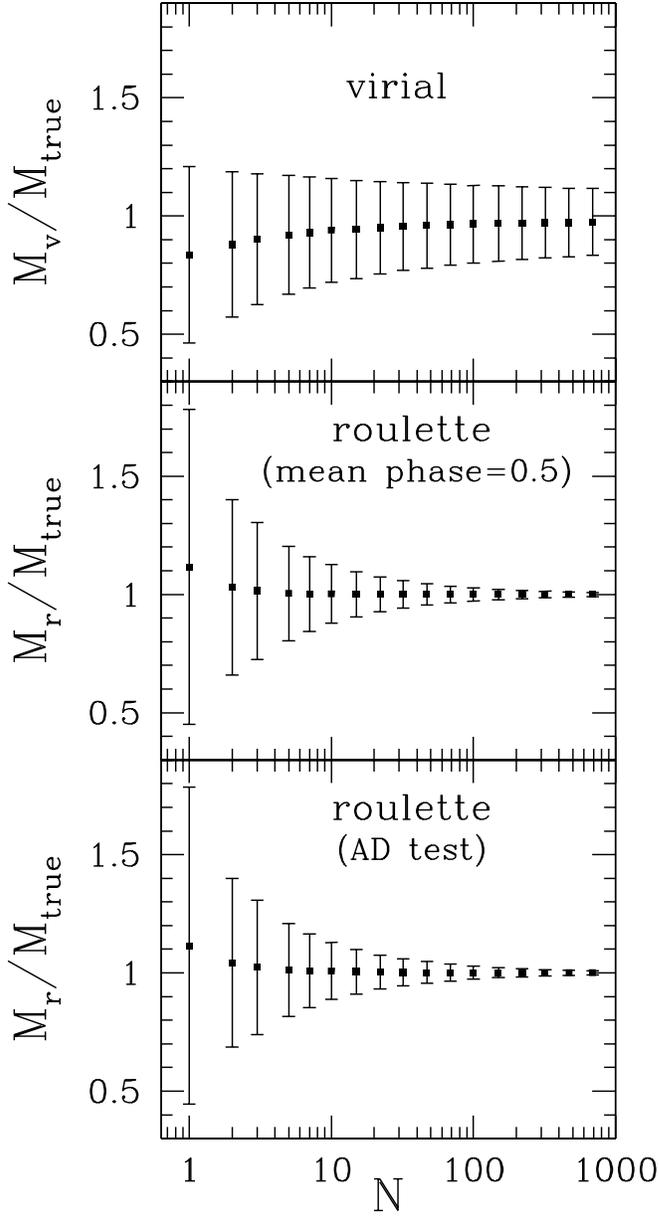}
\label{fig:r}
%\epsfxsize=17cm
%\epsfysize=17cm
%\epsfbox{f1.eps}
\end{center}
\caption{
Performance of the virial and roulette estimators tested with 
$N$ bodies on orbits with random eccentricities $0<e<1$ and
semi-major axes $0<a<\amax$. The mean value and standard 
deviation of the estimated mass are shown in the figure as
a function of $N$. Two versions of the
roulette estimator are shown: the mean-phase method and 
the Anderson-Darling (AD) test for the cumulative distribution of phases.
}
\end{figure}
%%%%%%%%%%%%%%%%%%%%%%%%%%%%%%%%%%%%%%%%%%%%%%%%%%%%%%%%%%%%%%
%%%%%%%%%%%%%%%%%%%%%%%%%%%%%%%%%%%%%%%%%%%%%%%%%%%%%%%%%%%%%%
\begin{figure}
\begin{center}
\plotone{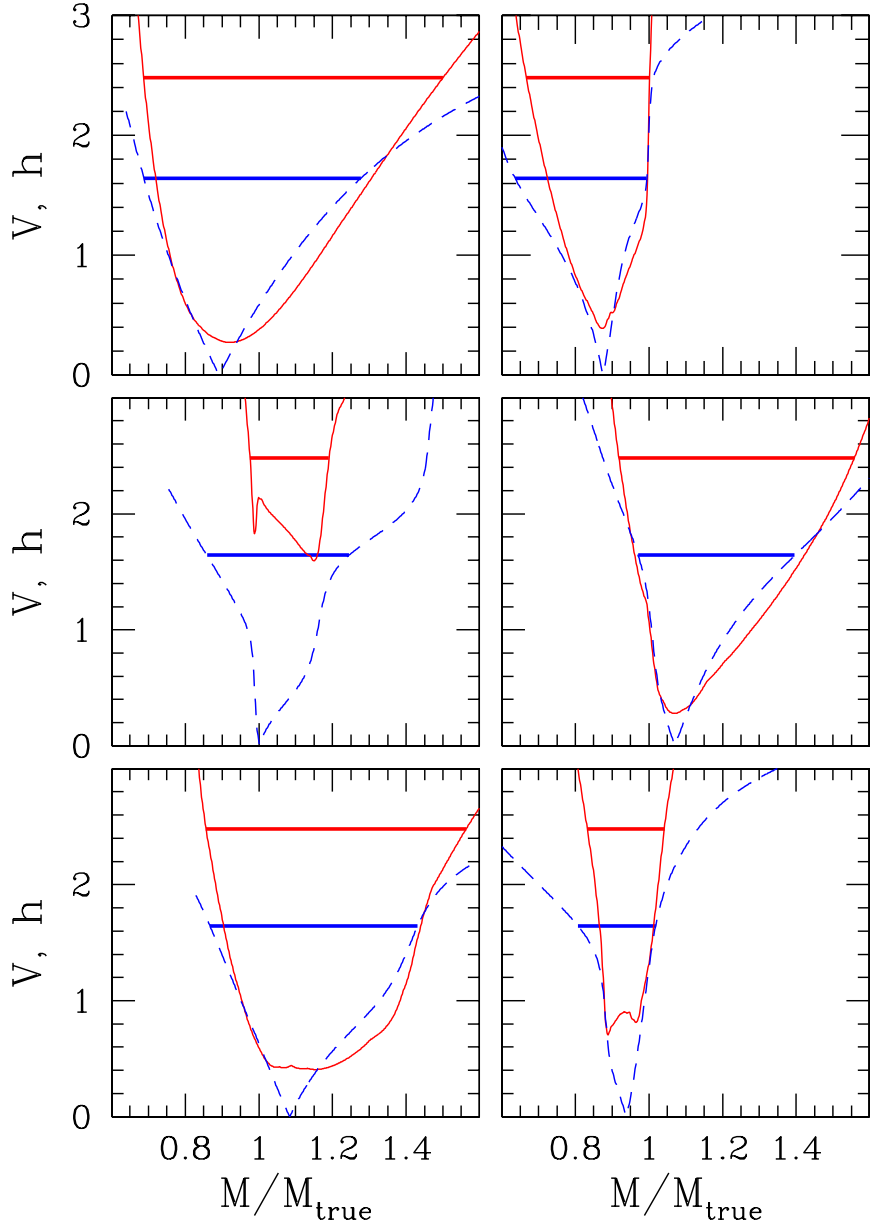}
\end{center}
\caption{
Roulette analysis of six randomly selected data sets with $N=10$
from the sample in Figure~4. Each panel displays the results for 
one data set. 
The measure $V$ of the cumulative-distribution estimator is found as a 
function of trial mass $M$ and shown by solid curve. The minimum of 
$V(M)$ gives the best-fit $M$, and $V=2.48$ defines the 90\% confidence 
interval for $M$ (shown by the horizontal line). The mean-phase estimator 
compares $\gav(M)$ with $1/2$, and the dashed curve shows 
$h=|\gav-0.5|(12N)^{1/2}$ as a function of trial $M$. Then the best-fit 
is where $h=0$, and the 90\% confidence interval is defined by $h=1.643$ 
($0.35<\bar{g}<0.65$).
}
\end{figure}
%%%%%%%%%%%%%%%%%%%%%%%%%%%%%%%%%%%%%%%%%%%%%%%%%%%%%%%%%%%%%%

%############################################################################

\section{The case of distributed mass}

Suppose we observe $N$ bodies moving in an unknown potential $\Phi({\bf r})$
which is created by a distributed mass with an unknown density profile. 
Hereafter for brevity 
we call the gravitating mass ``halo'' (keeping in mind a dark-matter 
halo of a galaxy) but it could also be a stellar cluster or a cluster 
of galaxies. We need only assume that the cluster has a sufficient 
number of members, so that gravity fluctuations due to motions in the 
cluster average out, a quasi-steady $\Phi({\bf r})$ is well defined,
and each observed body moves on a well-defined bound orbit in this 
potential. We now apply the roulette method to estimate $\Phi(r)$.

We limit our consideration to spherically symmetric potentials;
then an orbital motion is always confined to a plane.
Bound orbits in a potential $\Phi(r)$ are not closed in general,
however, they still have well-defined pericenter $\rp$ and apocenter $\rap$.
The bodies move periodically from the pericenter to the apocenter and 
back with a period $T$. This ``radial'' period is in general not equal 
to the period of the azimuthal motion. A typical example of such an orbit
is a rosette: a superposition of radial oscillations and azimuthal 
precession (e.g. Binney \& Tremaine, 1987).

Since $\rp$ and $\rap$ are well defined for each orbit, a moving body 
has a well defined orbital phase $g$: the time of motion from the current 
radius $r$ to the nearest pericenter (in the past or in the future)
divided by half period, $T/2$. If the body is observed at a random moment
of time then the phase should be a random number $0<g<1$ and we can apply 
the roulette test.

Any trial model $\Phi(r)$ can be tested as it gives certain phases $g_i$ 
to the observed bodies $i=1,...,N$. The model is good if $g_i$ are 
consistent with the Poisson distribution, which can be checked in 
exactly the same way as it was done for point-mass potentials in \S~3. 
In this way, the roulette test sorts out good models $\Phi(r)$ from bad ones. 

In practice, one deals with a parametrized family of potentials assuming a 
certain functional shape of $\Phi(r)$. Estimating $\Phi(r)$ is then reduced 
to constraining the allowed range of a small number of parameters.
The parametrization is consistent with the data if it passes the roulette 
test with some set of parameters. The test will signal if the assumed 
parametrization of $\Phi(r)$ is far from reality: the model will not be able 
to produce Poisson phases if $N$ is sufficiently large. 
For example, a point-mass model with one parameter $M$ will not give 
Poisson $g_i$ if the bodies actually move inside a homogeneous halo. 
Thus, the roulette method checks the assumed parametrization and 
simultaneously finds the best-fit parameters. In concrete problems, one 
may have an idea of the possible functional shape of the potential.
For example, dark-matter halos are expected to have certain shapes 
predicted by cosmological simulations of structure formation
(e.g. Navarro, Frenk, \& White 1996).

For the illustrative purpose, we consider below a simple potential with
two parameters $b$ and $m$,
\be
\label{eq:isoch}
  \Phi(r)=-\frac{Gm}{b+(r^2+b^2)^{1/2}}.
\ee
This is the well-known isochron potential (see e.g. Binney \& Tremaine 1987);
it describes a gravitating mass $m$ distributed in space so that most
of it resides inside the characteristic radius $b$. The corresponding
mass density is given by the Poisson equation ${\rm div}\,\Phi=4\pi\rho$.
At $r\ll b$, the density is $\rho\approx const=(3/16\pi)(m/b^3)$. 
At $r>b$ the density falls off quickly and $\Phi(r)$ tends to the 
point-mass potential $\Phi(r)=-Gm/r$. The equation of motion 
in the isochron potential can be integrated analytically and 
the orbit shape can be found (see Appendix~B).

\subsection{Phase-energy correlation}

Let us test the efficiency of the roulette estimator with Monte-Carlo 
generated sets of $N$ test bodies moving in an isochron potential.
The true potential has known parameters $\mtrue$ and $\btrue$. We assume in 
this test that the bodies have random orbital energy $E_0<E<0$ and random 
eccentricity $0<e<e_{max}$ defined in Appendix.\footnote{More precisely,
we assume in this example a Poisson probability distribution for 
$\sqrt{-E}$ between $0$ and $\sqrt{-E_0}$ and a Poisson distribution for 
$e$ between $0$ and $e_{max}$.}
The minimum $E_0=-(G\mtrue/2\btrue)=\Phi_{\rm true}(0)$ 
corresponds to a body at rest at $r=0$, and $e_{max}(E)=1-E/E_0$ is
the maximum eccentricity for an orbit with a given $E$. 

The created data set is studied by the roulette estimator which does not 
know $\mtrue$, $\btrue$. The estimator calculates the orbital phases 
$g_i(m,b)$ and compares them with the Poisson distribution using the 
Anderson-Darling measure $V$ as described in \S~3. Then it finds the best 
fit $(m_0,b_0)$ where $V(m,b)$ has a minimum.

We, however, know $\mtrue$ and $\btrue$ and can check the accuracy 
of the obtained estimate $(m_0,b_0)$. This check is made in Figure~6 
(upper panel) for 300 simulated data sets with $N=32$. We plot the results 
in the $q$-$m$ plane, where $q=m^{1/3}/b$ is a density parameter. 
The choice of $q$ and $m$ as two independent parameters (instead of $b$ 
and $m$) has a reason that will become clear below.

%%%%%%%%%%%%%%%%%%%%%%%%%%%%%%%%%%%%%%%%%%%%%%%%%%%%%%%%%%%%%%
\begin{figure}
\begin{center}
\plotone{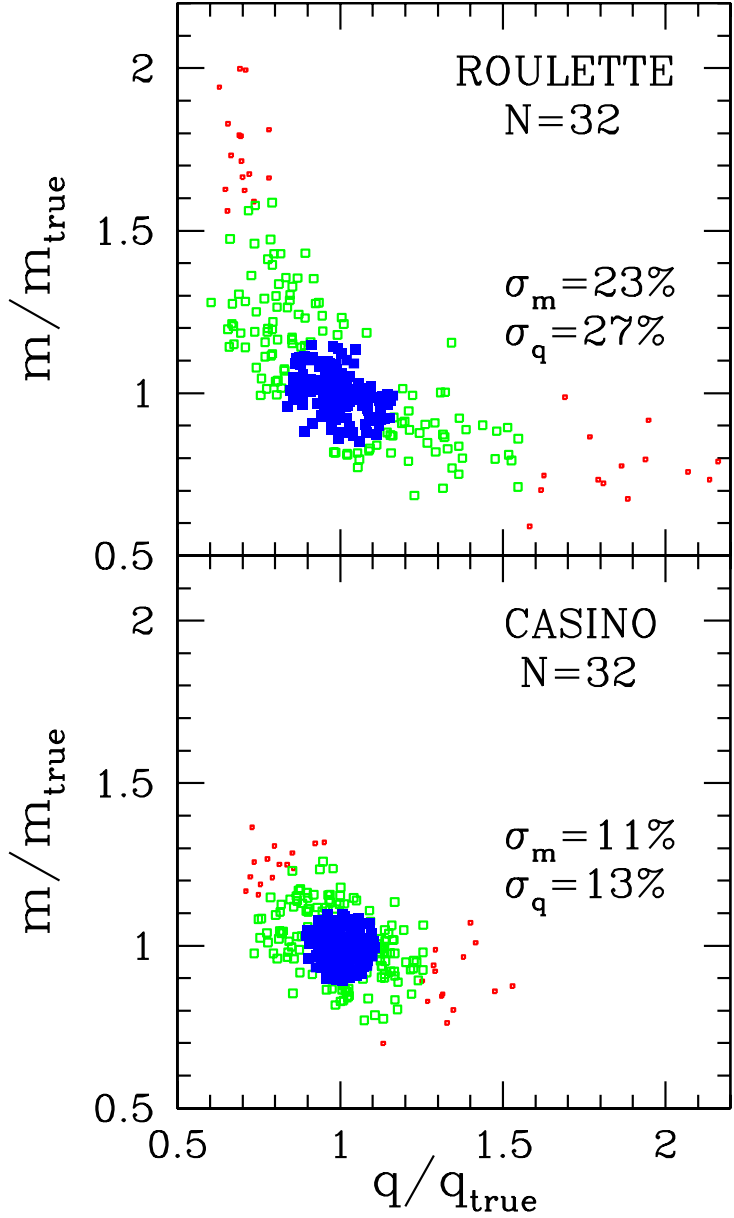}
\label{fig:bf}
%\epsfxsize=17cm
%\epsfysize=17cm
%\epsfbox{f1.eps}
\end{center}
\caption{300 best fits to the isochron potential probed with $N=32$ bodies 
randomly drawn from a population with random orbital energy and eccentricity 
(see the text). The fits are shown in the $q-m$ plane where $q=m^{1/3}/b$;
$m$ and $b$ are the mass and size of the isochron halo (eq.~\ref{eq:isoch}).
(a) Roulette method (AD test). (b) Casino method. 
10\% worst points are in red, 50\% best points are in blue, and 40\%
intermediate points are in green.
}
\end{figure}
%%%%%%%%%%%%%%%%%%%%%%%%%%%%%%%%%%%%%%%%%%%%%%%%%%%%%%%%%%%%%%

One can see relatively large deviations of $(q_0,m_0)$ from $(\qtrue,\mtrue)$.
Interestingly, the deviations are concentrated in an 
elongated region in the $q-m$ plane and their origin can be understood. 
First, we note that a wrong choice of $m$ affects mostly $g_i$ of bodies 
at $r>b$, and a wrong choice of $q$ affects $g_i$ of bodies at $r<b$.
Consider, for example, $m>\mtrue$. The orbital phases of the bodies at 
$r>b$ are then biased to the apocenter. This can be compensated 
if we choose $q<\qtrue$: then the bodies at $r<b$ will 
be biased to the pericenter. Thus, $m>\mtrue$ and $q<\qtrue$ can give 
the correct $\gav=1/2$ even when they deviate significantly from 
$\mtrue$ and $\qtrue$. The total distribution of $g_i$ is then 
biased to {\it both} apocenter and pericenter, and there is a lack 
of intermediate $g_i\sim 1/2$. 
At large $N$, the Anderson-Darling test notices such a non-Poisson
distribution of $g_i$ and rules out $(q,m)$ that are far from 
$(\qtrue,\mtrue)$. With increasing $N$, the acceptable $(q,m)$
converge to the true values. However, this convergence is relatively slow.

This problem is general for estimations of distributed-mass potentials 
that have more than one parameter: there are directions in the parameter
space where $\gav=1/2$ is satisfied and the non-flatness of the 
$g_i$ distribution has to be detected in order to rule out false 
parameters. The non-flatness is more difficult to detect than a deviation
of $\gav$ from 1/2 because of a lower signal-to-noise ratio, and 
therefore larger $N$ are required to estimate the potential accurately.

Fortunately, there is an efficient way to avoid this problem and
improve the accuracy of the roulette estimator. 
Indeed, we note that the large deviations of $(q_0,m_0)$ from 
$(\qtrue,\mtrue)$ are accompanied by a strong intrinsic correlation 
between $E_i$ and $g_i$. In our example above, $m_0>\mtrue$ and $q_0<\qtrue$,
the bodies on small orbits (small $E$) are found near the pericenter 
and the bodies on large orbits (large $E$) are near the apocenter. 
The orbital energy $E_i$ is an unknown integral of motion
(the other integral, angular momentum ${\bf l}_i={\bf r}_i\times{\bf v}_i$, 
is known from the data), and our observation can be generalized as 
follows: large deviations of the best-fit parameters from the true values
occur when the orbital phases and unknown integrals of motion are adjusted
in a special way and show an intrinsic correlation. Real, truly random, 
$g_i$ should show no correlation with integrals of motion.
Thus, the roulette method can be improved by tracking this correlation 
and making sure it is absent for the best fits.
In particular, in our halo problem, we should track the $E-g$ correlation.

A standard way to detect a monotonic (positive or negative) correlation 
is to calculate the cross-correlation coefficient 
$\sum (E_i-\bar{E})(g_i-\bar{g})$. We will use a better
measure that would signal any type of correlation between $g_i$ and $E_i$.
First we express the problem mathematically in a convenient way.
Given $g_i$ and $E_i$ we can sort the bodies in two ways: by $g$ and by 
$E$. Let us denote their phase and energy numbers by $i$ and $j$,
so that $g_{i+1}>g_i$ and $E_{j+1}>E_j$. The absence of correlation 
between $E$ and $g$ means that the permutation $i\rightarrow j$ is random. 
This permutation can be viewed as a mixing of $N$ cards (bodies), and 
in a fair card game we expect a truly random mixing. Plotted on the
$i-j$ plane, a random permutation $j(i)$ is represented by a set of $N$
points that should be consistent with a homogeneous 2D distribution in the 
square $(1,N)\times(1,N)$. Inhomogeneity would signal a correlation.

The consistency with homogeneity can be checked as follows.
Consider one point $i,j(i)$. It divides the square into four quadrants:
(1) $i^\prime<i$, $j^\prime<j$, 
(2) $i^\prime<i$, $j^\prime>j$, 
(3) $i^\prime>i$, $j^\prime<j$, 
(4) $i^\prime>i$, $j^\prime>j$. 
Denote the number of points that fall into each quadrant as 
$N_1,N_2,N_3,N_4$. For a random mixing $j(i)$ the mean expectation values
for these numbers are proportional to the areas of the corresponding 
quadrants. Only one of the four numbers is independent,
which we choose to be $N_1$ (we have $N_2=i-1-N_1$,
$N_3=j-1-N_1$, and $N_4=N-i-N_3=N-i-j+1+N_1$).
%We have the relations
%$N_1+N_2=i-1$, $N_3+N_4=N-i$ 
%$N_1+N_3=j-1$, $N_2+N_4=N-j$. 
Random mixings $j(i)$ give the hypergeometric probability distribution for 
$N_1$ with the mean expectation value and variance, 
\be
 <N_1>=(j-1)\frac{(i-1)}{(N-1)}, \qquad
 \Var(N_1)=\frac{(N-i)}{(N-2)}(i-1)\frac{(j-1)(N-j)}{(N-1)^2}.
\ee
$N_1$ is consistent with random mixing if its deviation 
from $<N_1>$ is comparable to $[\Var(N_1)]^{1/2}$.
The inconsistency is measured by
\be
   V_c[i,j(i)]=\frac{(N_1-<N_1>)^2}{\Var(N_1)}.
\ee
Finally we define a measure that should signal deviations from a
homogeneous distribution in any part of the square $(1,N)\times(1,N)$,
\be
  V_c=\sum_{i=2}^{N-1}V_c[i,j(i)].
\ee

$V_c$ plays the same role for the card test [randomness of mixing $j(i)$]
as the Anderson-Darling measure $V$ played for the roulette test 
(randomness of phase $g_i$). Using the Monte-Carlo technique we have 
calculated numerically the statistics of $V_c$ for truly random card 
mixings. We thus have the probability distribution of $V_c$, its mean 
expectation $<V_c>$, and variance $\Var(V_c)$. A mixing $j(i)$ with some 
$V_c$ will not pass our card test at a significance level $\xi$ if the 
probability that a fair $V_c^\prime>V_c$ equals $\xi$.

\subsection{Casino}

For the true potential $\Phi_{\rm true}(r)$ we should have random 
orbital phases $g_i$ {\it and} random energy-phase mixing $j(i)$. 
Thus we expect to deal with a fair casino, 
\begin{eqnarray}
 {\rm fair~casino} \equiv {\rm fair~roulette} + {\rm fair~cards}. 
\end{eqnarray}
For a false trial potentials $\Phi(r)$ we will detect that its casino is 
unfair by looking at the combination,
\be
\label{eq:chi}
  \chi^2(\Phi)=\frac{V^2}{\Var(V)}+\frac{V_c^2}{\Var(V_c)}.
\ee
Our best bet for $\Phi(r)$ is the potential that minimizes $\chi^2$.

A fair casino has a certain cumulative probability distribution 
$P(\chi^2)$ which we have calculated numerically. It is shown in Figure~7.
For example, one can see from the figure that a model that gives
$\chi^2\approx 16$ is good with 90\% confidence (10\% significance
of inconsistency with the null hypothesis of fair casino). 
Using $P(\chi^2)$ one can sort out acceptable potentials at any given
confidence level.

%%%%%%%%%%%%%%%%%%%%%%%%%%%%%%%%%%%%%%%%%%%%%%%%%%%%%%%%%%%%%%
\begin{figure}
\begin{center}
\plotone{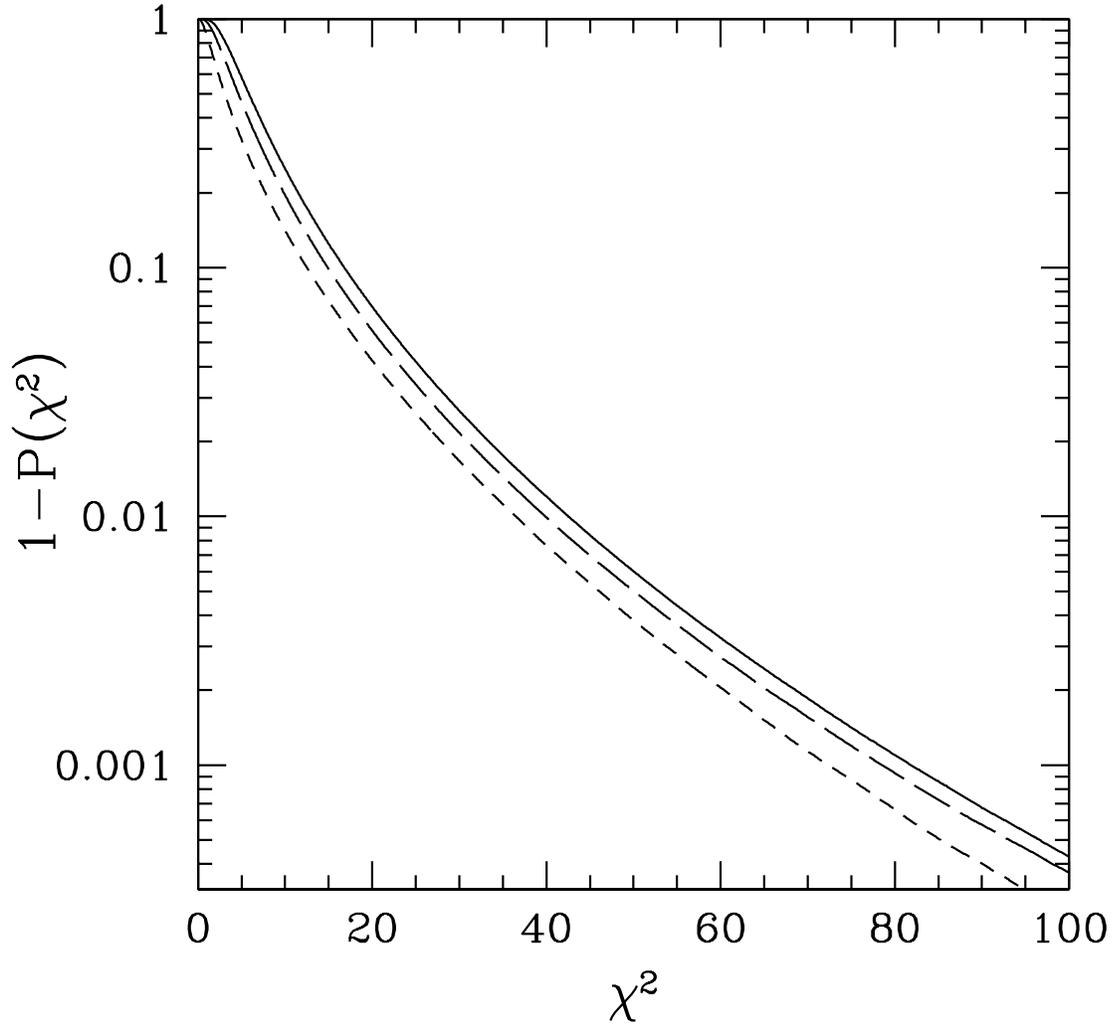}
\label{fig:cas}
%\epsfxsize=17cm
%\epsfysize=17cm
%\epsfbox{f1.eps}
\end{center}
\caption{Cumulative distribution of the casino measure $\chi^2$ 
defined in equation~(\ref{eq:chi}). Short-dashed, long-dashed, and
solid curves show the cases $N=10$, $32$, and $100$, respectively.
}
\end{figure}
%%%%%%%%%%%%%%%%%%%%%%%%%%%%%%%%%%%%%%%%%%%%%%%%%%%%%%%%%%%%%%

Let us illustrate the efficiency of the casino method by applying it to the 
isochron potential. We repeat our Monte-Carlo simulation shown in the upper 
panel of Figure~6, but now we get the best fit by minimizing $\chi^2$ rather 
than $V$. The results are shown in the lower panel of Figure~6. 
The accuracy of the casino estimator is improved significantly compared 
with the simple roulette. The standard deviations of the best fits from 
$(\qtrue,\mtrue)$ are now 11-13\%, and 90\% of the best fits show 
deviations less than $\sim 25\%$.

We think that the casino method extracts completely the available 
information from the data and does the best estimate of $\Phi$ that could
possibly be done. As an illustration, we show in Figure~8
the results of the method application to three 
Monte-Carlo-generated data sets with N=100. In a similar way, it would 
be applied to real data. The only difference is that here we know the true
potential (and the method does not). The results are presented as 90\% 
confidence area on the $q-m$ plane. The data set analyzed in panel (a) is 
drawn from a population with random orbital energy $E$ between $0$ and 
$E_0=-(G\mtrue/2\btrue)$ and random $e$. It gives a compact
confidence area and both $q$ and $m$ are correctly measured with
accuracy better than 10\%. Panel (b) shows the result for a data
set with $E$ randomly drawn from the interval $E_0<E<(19/20)E_0$.
In this case, the bodies sit well inside $b$ and the method cannot say 
anything about the total mass of the halo $m$. This is evident for us because
we know $\mtrue$ and $\btrue$, so we know that the bodies are inside $b$.
However, the method initially did not know that. It finds that it cannot 
say anything about $m$ and hence {\it finds} that the observed bodies 
move in a central part of a distributed mass. It measures well 
(with 4\% accuracy) all it could possibly measure: the central density 
$\rho=(3/16\pi)q^3$. 
Panel (c) shows the case where $E$ is drawn from the interval
$E_0/20 < E < 0$. Here the dominant majority of the bodies
are far outside $b$ and the method measures well the total mass $m$
but cannot say anything about $b$ and $\rho$.
Thus it finds that the test bodies move in essentially a point-mass
potential and estimates this mass with accuracy better than 10\%.

%%%%%%%%%%%%%%%%%%%%%%%%%%%%%%%%%%%%%%%%%%%%%%%%%%%%%%%%%%%%%%
\begin{figure}
\begin{center}
\plotone{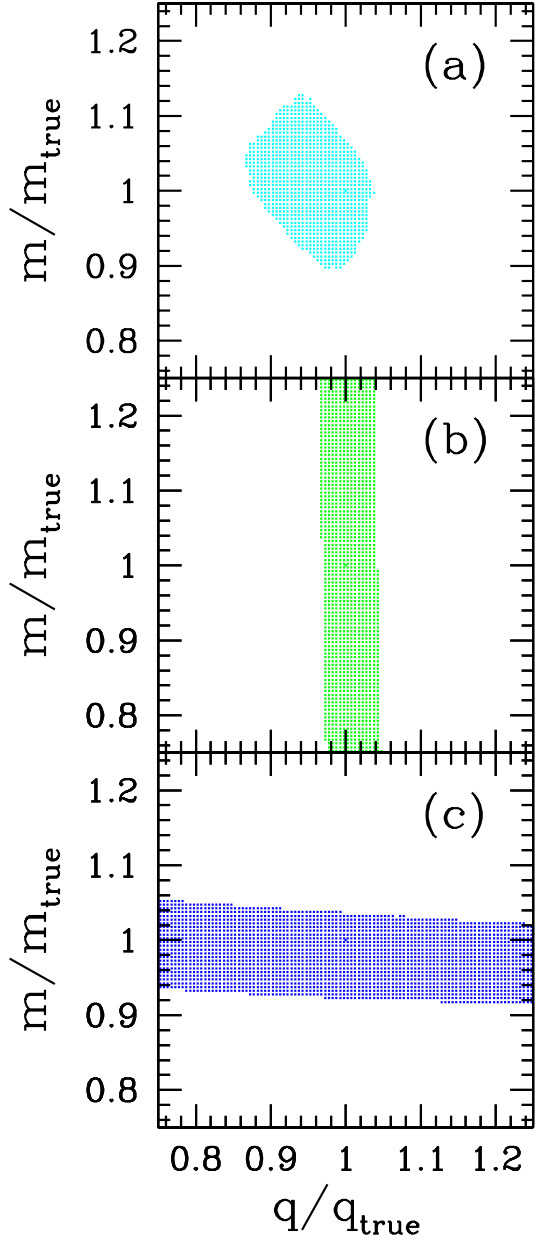}
\label{fig:illustr}
%\epsfxsize=17cm
%\epsfysize=17cm
%\epsfbox{f1.eps}
\end{center}
\caption{ 90\% confidence area on the $q-m$ plane obtained with the 
casino method for a Monte-Carlo generated data set of $N=100$ bodies 
with random $E$ and $e$. (a) $E$ is drawn from the interval $E_0<E<0$.
(b) $E$ is drawn from $E_0<E<(19/20)E_0$ (small orbits inside the halo).
(c) $E$ is drawn from $(E_0/20)<E<0$ (large orbits outside the halo).
}
\end{figure}
%%%%%%%%%%%%%%%%%%%%%%%%%%%%%%%%%%%%%%%%%%%%%%%%%%%%%%%%%%%%%%

We have also checked the method ability to reject incorrect parametrizations
of $\Phi(r)$. For example, if the test bodies move inside a homogeneous halo, 
$N=10-15$ is enough to reject the point-mass model: the best-fit then has too 
high $\chi^2$ and does not pass the casino test. The confidence level of 
rejection increases quickly with $N$.

%############################################################################

\section{Discussion}

\subsection{Limitations of the roulette method}

There are two situations where the roulette method should be applied 
with caution. 

\noindent
1. ---- Orbital phases of the test bodies are not independent.
This happens if, for example, the satellites are in orbital resonances 
with each other and the phase-locking effect takes place, 
as is the case for planets in the outer solar system. 

\noindent
2. ---- 
The snapshot of the test bodies has a finite size comparable to the 
size of their orbits. Then there is a risk that the orbits 
extend beyond the snapshot boundary, and some bodies are observed 
preferentially near their pericenters just because they would not be seen 
otherwise. This bias to the pericenter can be significant only for 
satellites with highly eccentric orbits and it can be corrected:
for each orbit reconstructed with a trial potential, one can check 
whether it extends beyond the snapshot boundary and calculate the fraction 
of the orbital period that the body spends beyond the boundary.

\subsection{Applications}

We briefly discuss below three astrophysical situations where
the proposed method can be applied in its full 3D version
developed in this paper.\footnote{We are grateful to Scott Tremaine who
pointed out the applications outlined in \S\S~5.2.1 and 5.2.3.}

\subsubsection{The Oort limit}

The mass content of the Galactic disk in the solar neighborhood
is inferred from the vertical motions of the nearby stars with respect 
to the galactic plane (see, e.g., chapt. 4.2 in Binney \& Tremaine 1987).
Mathematically, the problem reduces to a reconstruction of a
one-dimensional potential $\Phi(z)$ from a snapshot of 
one-dimensional motions of $N$ test bodies.

The Jeans equation was previously used to reconstruct $\Phi(z)$, which is 
not the most efficient method because it involves (arbitrary) binning of 
the data and numerical triple differentiation.
Orbital roulette, in its full casino version, should be able to use the 
data optimally. It is expected to find $\Phi(z)$ that gives the 
stars random phases of vertical motion (roulette test), and the phases 
should be uncorrelated with the stars' orbital energies (card test).
These conditions are necessary and sufficient for $\Phi(z)$ to be 
realistic.

\subsubsection{The mass of Sgr A*}

Observations of stellar motions in the Galactic Center allow one to estimate 
the mass of the central dark object Sgr~A*, which is believed to be a giant 
black hole (Genzel et.~al. 2000; Sch\"odel et.~al. 2002, Ghez et.~al. 2003). 
Normally, only sky-projected positions of stars are available, with the
exception of a few tightly bound stars whose orbits have been partially 
mapped out.

By analyzing the data of Genzel et.~al. (2000) we have recently found
that the young stellar population in the Galactic center forms a thin disk 
(Levin \& Beloborodov 2003). The inferred orientation of the disk plane was 
confirmed by further analysis of Genzel et.~al. (2003).
This finding gives a full 3D information for instantaneous motions of ten 
disk stars, which can be used to estimate the gravitational potential.
We have applied the orbital roulette to these data and obtained 
a new independent estimate of the black hole mass (in preparation).

\subsubsection{Galactic potential probed by SIM}

Observations of satellites of our Galaxy constrain 
its gravitational potential and mass content.
Within a decade, the Space Interferometry Mission (SIM)
%see http://sim.jpl.nasa.gov/whatis/) 
can provide the full 3-D information on the motion of tens of the satellites, 
and the data should be used most efficiently.

The most advanced method to date, which is based on the Bayesian analysis, 
was developed by Little \& Tremaine (1987) and Kochanek (1996).
Little \& Tremaine (1987) assumed that the satellites' velocities are 
either radial or have isotropic distribution while Kochanek (1996) assumed 
some {\it \`a priori} form of the satellites' distribution function. 
The orbital roulette, combined with the 3-D data, will alleviate the need 
for {\it \`a priori} assumptions. In the previous section we have 
demonstrated how our method estimates the mass and size of a spherically 
symmetric halo. A similar analysis can be done for the SIM data with a 
realistic model of the Galactic potential.

\acknowledgments
We thank Scott Tremaine for many helpful discussions. We also thank
Mark Wilkinson, the referee, for his suggestions that helped improve 
the presentation of the paper. 
Both authors acknowledge financial support from NSERC.
A.M.B. was supported by Alfred P. Sloan Foundation.

%##############################################################################

\section*{Appendix A: Keplerian orbits}

A Keplerian orbit around point mass $M$ is characterized by
two integrals of motion: orbital energy and angular momentum,
\be
\label{eq:E}
  E=\frac{v^2}{2}-\frac{GM}{r}, \qquad {\bf l}={\bf r}\times {\bf v}.
\ee
The semi-major axis $a$ and eccentricity $e$ of the orbit are given by 
\be
\label{eq:ecc}
   a=\frac{GM}{2|E|}, \qquad  1-e^2=\frac{2|E|l^2}{G^2M^2}.
\ee
The pericenter and apocenter radii are
\be
\label{eq:rp}
  \rp=a(1-e), \qquad \rap=a(1+e).
\ee
For the orbital roulette we need to determine the time of motion
from $\rp$ to a given $r$. It is convenient to calculate the time using
the relation
\be
 \dd t=\frac{\dd r}{v_r}=\left(\frac{2}{|E|}\right)^{1/2}r\,\dd\psi, 
\ee
where $\psi$ is defined by
\be
\label{eq:psi}
 \sin^2\psi\equiv\frac{|E|(r-\rp)}{GMe},
\ee
so that $r=\rp$ at $\psi=0$ and $r=\rap$ at $\psi=\pi/2$.
The full orbital period corresponds to $\psi$ changing from $0$ to $\pi$. 
The time of motion from a given $r$ to $\rp$ is
\be
\label{eq:trp}
  t_p(r)=\frac{T}{\pi}\left(\psi-e\frac{\sin 2\psi}{2}\right),
\ee
where
\be
  T=2\,t(\rp\rightarrow\rap)=\frac{\pi GM}{\sqrt{2}|E|^{3/2}}.
\ee
is the orbital period.
Time average of a magnitude $Z$ over the orbit is given by
\be
  <Z>=\frac{1}{T}\int_0^T Z\dd t=\frac{2}{T}
      \left(\frac{2}{|E|}\right)^{1/2}\int_0^{\pi/2}Z r \dd\psi.
\ee

Suppose an observed body has a measured position ${\bf r}$ and 
velocity ${\bf v}$, and denote the angle between ${\bf r}$ and ${\bf v}$
by $\alpha$. If we do not know the central mass $M$, the orbital parameters
of the body and its phase $g=t_p(r)(T/2)^{-1}$ will depend on the 
assumed $M$ or $x=GM/v^2r$. In particular, we have
\be
   e(M)=\left[1-\frac{(2x-1)}{x^2}\sin^2\alpha\right]^{1/2},
\ee 
\be
  \sin^2\psi(M)=\frac{x-1-xe}{2xe},
\ee
\be
  g(M)=\frac{2}{\pi}\left(\psi-\frac{e}{2}\sin 2\psi\right),
\ee
\be
  \frac{\dd g}{\dd M}=\frac{1}{\pi eMx}
 \left[\frac{(1-e\cos 2\psi)}{\sin 2\psi}\left[1-\left(1-\frac{1}{x}\right)^2
      \frac{\sin^2\alpha}{e^2}\right]-\left(1-\frac{1}{x}\right)\sin^2\alpha
       \,\sin 2\psi\right].
\ee
These expressions are used in section 3.5.

%##############################################################################

\section*{Appendix B: Isochron orbits}

In contrast to the Keplerian case, we now have a characteristic
scale $b$. It is convenient to use a dimensionless variable $s$
instead of $r$,
\be
  s=1+\sqrt{1+\frac{r^2}{b^2}}, \qquad r=b\sqrt{s(s-2)}.
\ee
The pericenter and apocenter radii are derived from the equation,
\be
   \frac{v_r^2}{2}=E-\Phi-\frac{l^2}{2r^2}=0,
\ee
which gives
\be
  s_{1,2}=1+\frac{E_0}{E}\mp\left[\left(\frac{E_0}{E}-1\right)^2
          +\frac{l^2}{2b^2E}\right]^{1/2}.
\ee
Here $E_0=\Phi(0)=-(Gm/2b)$ is the minimum possible energy for an 
orbit in the isochron potential. It is convenient to define a 
``semi-major'' axis $a$ in the $s$-space as
\be
  2a=b(\sp+\sap-2)=\frac{Gm}{|E|}.
\ee
The relation between $a$ and $E$ is exactly the same as in the 
Keplerian case, and $a$ has a normal Keplerian meaning at $a\gg b$.
At $r<b$, however, the $s$-space $a$ is very different; it can 
never be smaller than $b$ and can be written as 
\be
   a=b\frac{E_0}{E} >b.
\ee

Using $\dd t=\dd r/v_r$, one finds
\be
  \dd t=\frac{b}{2|E|}\frac{(s-1)\dd s}{\sqrt{(\sap-s)(s-\sp)}}.
\ee
Let us define angle $\psi$ by 
\be
   \sin^2\psi=\frac{s-s_1}{s_2-s_1}.
\ee
Then the time of motion from a given $r$ to the pericenter $r_1$ is
\be
\label{eq:trp_s}
  t_p(r)=\frac{T}{\pi}\left(\psi-e\frac{\sin 2\psi}{2}\right),
\ee
where $T$ is the period of radial oscillations, 
\be
  T=2\,t(\rp\rightarrow\rap)=\frac{\pi Gm}{\sqrt{2}|E|^{3/2}},
\ee
and $e$ is the ``s-space eccentricity'' defined by
\be
  e=\frac{b(\sap-\sp)}{2a}.
\ee 
The relation between $E$ and $T$ is the same as in 
the Keplerian case, and the formula for $t_p$ is formally the 
same (although the eccentricity $e$ is defined in the $s$-space here).

The eccentricity can also be expressed in terms of $E$ and $l$,
\be
   e^2=\left(1-\frac{E}{E_0}\right)^2+\frac{l^2E}{2b^2E_0^2}.
\ee
Linear orbits with $l=0$ have a maximum eccentricity 
\be
  e_{\rm max}(E)=1-\frac{E}{E_0}.
\ee

In the limit $r\ll b$ (which corresponds to $\tilde{E}=E-E_0\ll |E_0|$) 
the orbital motion occurs inside the homogeneous spherically-symmetric 
halo with density 
\be
  \rho=\frac{3m}{16\pi b^3}.
\ee

%########################################################################


\newpage

\begin{references}
\reference{} Anderson, T.~W., \& Darling, D.~A.~1952, Ann.~Math.~Stat., 23, 193
\reference{} Bahcall, J.~N., \& Tremaine, S.~1981, ApJ, 244, 805
\reference{} Binney, J., \& Tremaine, S.~1987, {\it Galactic Dynamics}, 
             Princeton University Press, New Jersey
\reference{} Genzel, R., Pichon, C., Eckart, A., Gerhard, O.~E., \& Ott, T.
             2000, MNRAS, 317, 348
\reference{} Genzel, R., et.~al.~2003, ApJ, 594, 812
\reference{} Ghez, A.~M., et.~al.~2003, ApJ, 586, 127
\reference{} Kochanek, C.~S.~1996, ApJ, 466, 638 
\reference{} Kolmogorov, A.~1941, Ann.~Math.~Stat., 12, 461 
\reference{} Levin, Y., \& Beloborodov, A.~M.~2003, ApJ, 590, L33
\reference{} Little, B., \& Tremaine, S.~1987, ApJ, 320, 493
\reference{} Navarro, J.~F., Frenk, C.~S., White, S.~D.~M. 1996, ApJ, 462, 
             563
\reference{} Page, T.~1952, ApJ, 116, 63
\reference{} Press, W.~H., Flannery, B.~P., Teukolsky, S., \& Vetterling, W. 
             1992, Numerical Recipes in C: The Art of Scientific 
             Computing (Cambridge: Cambridge Univ. Press)
\reference{} Sch\"odel, R., et.~al. 2002, Nature, 419, 694
\end{references}
\end{document}